\documentclass[pre,aps,nofootinbib,eqsecnum]{revtex4}
\usepackage{setspace}

\usepackage{amsmath,amssymb,bm}
\usepackage[colorlinks]{hyperref}
\usepackage[all]{hypcap}
\usepackage{graphicx,psfrag,float,epsfig}
\usepackage{mathrsfs,wasysym}
\usepackage{epstopdf}
\usepackage{comment}
\usepackage{pbox}
\usepackage{array}
\usepackage{xspace}
\usepackage[usenames,dvipsnames]{color}

\DeclareMathAlphabet\mathbfcal{OMS}{cmsy}{b}{n}

\DeclareSymbolFontAlphabet{\mathrsfs}{rsfs}
\DeclareMathAlphabet{\mathcal}{OMS}{cmsy}{m}{n}

\DeclareSymbolFont{bbold}{U}{bbold}{m}{n}
\DeclareSymbolFontAlphabet{\mathbbold}{bbold}

\newcommand{\be}{\begin{equation}}
\newcommand{\ee}{\end{equation}}
\newcommand{\nn}{\nonumber}

\newcommand{\beq}{\begin{equation}}
\newcommand{\eeq}{\end{equation}}
\newcommand{\bea}{\begin{eqnarray}}
\newcommand{\eea}{\end{eqnarray}}

\def\ba{ {\bm a} }

\def\br{ {\bm r} }

\def\bv{ {\bm v} }

\def\calO{ {\cal O} }

\providecommand{\cg}[1]{{\textcolor{blue}{{CG: #1}}}\xspace}

\begin{document}

%%%%%%%%%%%%%%%%%%%%%%%%%%%%%%%%%

\title{Deriving Analtyic Solutions for Compact Binary Inspirals Without Recourse to Adiabatic Approximations }
\author{Chad R.~Galley}
\affiliation{	Theoretical Astrophysics, Walter Burke Institute for Theoretical Physics, California Institute of Technology, Pasadena, California 91125, USA}
\author{Ira Z.~Rothstein}
\affiliation{Department of Physics, Carnegie Mellon University,
	Pittsburgh, PA 15213, USA}

\vspace{0.3cm}

%%%%%%%%%%%%%%%%%%%%%%%%%%%%%%%%%%%%%%%%%%

\begin{abstract}
	We utilize the dynamical renormalization group formalism to calculate the real space
	trajectory of a compact binary inspiral for long times via a systematic resummation of secularly growing terms.
	This method  generates closed form solutions without orbit averaging, and the accuracy  can be systematically improved.
	The expansion parameter is $v^5 \nu \Omega(t-t_0)$ where $t_0$ is the initial time, $t$ is the time elapsed, and $\Omega$ and $v$ are the angular orbital frequency and initial speed, respectively. $\nu$ is the binary's symmetric mass ratio.
	We demonstrate
	how to apply the renormalization group method to resum solutions beyond leading order in two ways. First, we calculate the second
	order corrections of the leading radiation reaction force, which involves
	highly non-trivial checks of the formalism (i.e. its renormalizability). Second, we show how to systematically
	include post-Newtonian corrections to the radiation reaction force.
	By avoiding orbit averaging we gain predictive power and eliminate ambiguities in the initial conditions.
	Finally, we discuss how this methodology can be used to find
	analytic solutions to the spin equations of motion that are valid over long times.
\end{abstract}

\maketitle

%================================
\section{Introduction}
\label{sec:intro}
%================================

The recent detections of gravitational waves from a binary black hole coalescence~\cite{firstGWdetection, secondGWdetection} provide the first measurements of the dynamics of compact binaries and strong gravitational fields. 
To date the measured events involved relatively large black hole masses. 
For the first detection, several gravitational wave cycles were observed in LIGO's frequency band corresponding to a handful of orbits before merger. 
If compact binaries with lower total masses are observed then these sources will evolve for much longer times within the detector's bandwidth and will be amenable to analytic calculations via the post-Newtonian (PN) expansion.

In order to extract the most information from such long waveforms requires using templates that have been computed with the highest possible accuracy. 
The lengthy inspiral regime of the binary's evolution is found by solving the post-Newtonian equations of motion. 
A challenge to this program is the calculation of the radiation reaction forces, the leading piece of which starts at 2.5PN~\cite{BurkeThorne:Relativity, Burke:JMathPhys12}. 
The PN corrections at 1PN beyond leading order were calculated in \cite{IyerWill:PRL70, IyerWill:PRD52,Blanchet:1993ng, Galley:2012qs}. 
At 1.5PN there is a contribution from the ``tail effect", which was calculated in \cite{Blanchet:1987wq, Galley:2015kus}.
Higher order corrections have yet to be calculated, however, the leading contributions from spin-orbit and spin-spin effects are known~\cite{Will:2005sn, Wang:2007ntb}, which first appear at 1.5PN and 2PN orders, respectively, beyond the leading radiation reaction force.  
As such, solving the equations of motion, even numerically, would lead to errors which are of order the first unknown radiation reaction contribution.
This would bound the accuracy of the prediction given that the conservative pieces of the equations of motion are known to higher order.
To avoid this issue one can utilize the power loss to account for radiation reaction by using the adiabatic approximation (more on this below) which requires one to average over the orbit thereby generating information loss.
 
In addition to this aforementioned limitation, solving for the binary's orbital motion is often achieved numerically because there are nine (once we include spin) nonlinearly coupled, ordinary differential equations that need to be solved accurately in order to follow the orbit's inspiral. 
The orbit needs to be sampled at a sufficiently high rate so that the corresponding waveform is  sampled appropriately. 
These accuracy requirements and/or high sample rate are often a computational bottleneck for gravitational wave data analysis applications that require many waveforms to be generated and hence many numerical solutions of the PN equations of motion. 
Such applications include template generation for gravitational wave searches and parameter estimation studies using Markov-Chain Monte-Carlo algorithms.

Recent developments in implementing a ``precession-averaging'' procedure~\cite{Kesden:2014sla, Gerosa:2015tea} to the equations of motion, which utilize the separation of time scales present in precessing binaries helps, alleviate some of the computational pressures mentioned above, depending on the specific application. 
However, accurate, globally valid, analytical approximate solutions for the binary's evolution would certainly remove any such computational bottleneck while simultaneously providing useful analytical expressions for studying the complicated physics of precession dynamics (such as recent evidence for precessional instability~\cite{Gerosa:2015hba}).

Currently, there are two standard methods used to provide analytic solutions for the restricted case of compact binaries with component masses that are spinning but non-precessing.
The first is the ``adiabatic approximation'' and is based on equating the time-averaged flux of gravitational waves to the mechanical power lost by the binary (see, e.g.,~\cite{Maggiore, Blanchet:LRR}). 
This method is often used to calculate the approximate gravitational wave phase of the $(\ell, m) = (2,2)$ spherical harmonic mode in the ``restricted approximation'' where the waveform is constructed from the orbital phase up to a definite order in the PN expansion, but the amplitude is taken to be the leading order quantity with no PN corrections. 
The second is the ``improved variation of constants method''~\cite{damour1983gravitational}. 
Here, the idea is to assume that the integration constants of the conservative part of the binary's dynamics exhibit a long-time evolution relative to the orbital period.
Equations for these integration ``constants'' are then found using the method of variation of constants, which are then solved. 
This approach is also known as the ``method of osculating orbits'' and is very closely related to ``multiple scale analysis.''
Recent work in~\cite{Chatziioannou:2016ezg} builds off of~\cite{Kesden:2014sla} and uses multiple scale analysis to find accurate analytic approximations for the orbit and the frequency-domain waveform of a precessing binary inspiral.

These approaches are based on averaging the PN equations of motion over the orbital period to help simplify the differential equations being solved. 
However, there are potential shortcomings with averaging that have been raised in~\cite{pound2008osculating, pound2005limitations}. 
In particular, the initial conditions used to solve the PN equations of motion and the orbit-averaged version are not the same so that comparing the resulting two solutions can be ambiguous.
In addition, there is an ambiguity in the period to use for the averaging procedure for eccentric orbits because there are different ways to characterize the time-scale of the orbit (such as the orbital period, eccentric anomaly, true anomaly, and mean anomaly). 
Using a different measure for the averaging can lead to different predictions over sufficiently long times.

Our work introduces a formalism that allows for the systematic solution of the PN equations of motion for a binary inspiral including radiation reaction forces and spin effects to an arbitrary order in the PN expansion. 
The method is based on applying ideas from renormalization group theory.
We do not implement any averaging procedures so that our solutions describe the binary's real-time orbital configuration at every instant of time.
The approach starts with a background (e.g., circular or eccentric) orbit and treats the radiation reaction force as a perturbation.
These perturbations grow secularly with time but can be resummed using the
 {\it Dynamical Renormalization Group} (DRG) method~\cite{Chen:1995ena}.
 The DRG method subsumes several approaches for the global analysis of differential equations~\cite{BenderOrszag}, including multiple scale analysis, boundary layer theory, and the WKB method. 
See~\cite{Chen:1995ena} and the appendix of \cite{Boyanovsky:2003ui} for pedagogical examples using the DRG method.
The DRG resummation is not to be confused with Pad\'e ``resummation'', which attempts to improve the 
radius of convergence of a perturbative expression. Pad\'e ``resummation'' is not a systematic
expansion as it does not resum any leading order pieces in a systematic way.  Different Pad\'e approximants 
often give different predictions at the same scale. Conversely, DRG literally resums higher order terms in 
the perturbation theory with a consistent power counting. 

Our focus here is to present the DRG method and demonstrate the internal consistency of the approach for non-spinning compact binary inspirals via higher order perturbative calculations. 
We will show how to find closed form solutions for the inspiral without 
recourse to the adiabatic approximation or orbit averages.
In a future paper, we will incorporate spin effects to obtain accurate and globally valid, real-time approximate solutions for the generic case of precessing compact binaries.

%================================
\section{Defining the systematics}
\label{sec:dsho}
%================================

We will perturbatively solve the PN equations of motion for compact binary inspirals, which are derived in an expansion where the binary's relative speed $v$ is small compared to the speed of light.
However, there is another power counting parameter for inspirals, namely,
\begin{align}
	\varepsilon \equiv v^5 \nu \Omega (t-t_0).
\end{align}
Here $t$ is the time elapsed since the initial time $t_0$, $\Omega$ is the initial angular orbital frequency, $v$ is the initial orbital speed, and $\nu$ is the binary's symmetric mass ratio.
The parameter $\varepsilon$ arises as a consequence of the secular growth due to the radiation reaction force  which is treated as a perturbation of the circular solution (or, more generally, energy-conserving motions) to the equations of motion. 
By performing a resummation, the accuracy of the perturbative solutions will be extended to much later times even when $\varepsilon$ is of order one.
Without such a resummation, the perturbative solutions would be of minimal utility.  
By resumming powers of $\varepsilon$ we are able to make precision predictions with 
well-defined systematics such that the result for the orbit is valid until the PN expansion
breaks down as the plunge is approached.

Our formalism allows us to go to arbitrary order in  $\varepsilon$ and allows for the systematic inclusion of PN corrections.
In this paper, we will demonstrate how to work to second order in $\varepsilon$.
If there were no higher order PN corrections then the resulting resummed solution is
valid up to times when
\beq
v^{10} \nu^2 \Omega (t-t_0) \sim  1. 
\eeq
However, in reality we must consider PN corrections that would contribute at lower orders.
% \cg{These next couple of sentences were confusing to me and didn't seem to mesh together.}
%However, in reality we must consider PN corrections that contribute at orders well below the 2.5PN order of the leading order radiation reaction force.
%\cg{[The previous sentence seems to refer to potential corrections but the next sentence seems to refer to radiation reaction corrections.]}
%We will demonstrate how to include these contributions by calculating the 1PN correction to the radiation reaction force. 
%If we ignored other sources of 1PN corrections then this result would be valid up to times when $v^8 \nu \Omega(t-t_0) \sim 1$.
We will also demonstrate how to include contributions from the radiation reaction force that are at  higher PN orders by calculating the 1PN correction to the orbital motion. 
It is important to realize that none of the results in this paper include all of the effects at a given order because our purpose is to present the method here. 
If we wished to perform the calculation including all 2PN effects, for instance, we would need to include the conservative potential up to 2PN, which is of course known, but we would also need to include the 2PN correction to the radiation reaction force which is presently unknown.

%================================
\section{Review of Renormalization Group Methodology}
%================================

For completeness, we present a lightning review of the logic behind the renormalization group (RG).
The DRG applies the logic of the RG to differential equations but the basic idea is the same.
Canonical RG applications are formulated within the context of a Lagrangian which will not
be the case for the DRG, though it is a simple exercise to embed the DRG into a Lagrangian 
formalism (necessarily for generic non-conservative systems~\cite{Galley:2012hx, Galley:2014wla}). However, doing so does not lead to any new insights (that we can see, at least). Thus, we will eschew such a treatment.

The basic algorithm is given as follows:
\begin{enumerate}
	\item Write down a background solution around which to perturb.  This solution is written in
	terms of ``bare" parameters (i.e., $A_B(t_0)$). These parameters implicitly depend upon the initial
	time $t_0$, away from which we flow.
	\item Use this background to calculate perturbatively the first correction to the equations of motion.
	The perturbation will in general have secular ``divergences,'' that is, terms that grow as $(t-t_0)$.
	\item Take this solution and write the bare parameters as renormalized parameters (i.e., $A_R(\tau)$) plus
	``counter-terms". These counter-terms will be proportional to $(\tau-t_0)$ and are chosen
	to eliminate the $t_0$ dependence of the aforementioned solution. $\tau$ is known as the
	subtraction point.  This step yields the ``renormalized'' solution.
	\item The renormalized solution must be independent of the choice of subtraction point $\tau$.
	The explicit dependence on $\tau$ in the solution  is cancelled by the implicit dependence of the renormalized parameters on $\tau$.
	One then uses this fact to derive a first-order differential equation (called the RG equation) for the renormalized parameter. 
	The right-hand side of the RG equation is called the ``beta function.''
	\item Solve the  RG equation for the parameter and choose $\tau=t$, the observation time.
	In so doing, all of the secularly growing terms are resummed at this order. 
\end{enumerate}
	
The ability to absorb divergences into the initial data, in the context of DRG,
is called ``renormalizability."  The renormalizabilty of the theory can be put on
firmer mathematical ground using  envelope theory as discussed in
\cite{Kunihiro:1995zt}. The basic notion is that a perturbative solution defines a family
of curves parameterized by $t_0$. Each of these solutions is only valid locally for times near $t_0$.
A global solution is then found by determining the envelope of this set of curves,
which is defined as the curve whose intersection with each curve in the family
is tangent to the given curve. 
	
The connection between the RG and global analysis is also manifest in holography.
Solving the equations of motion for a scalar field in anti-deSitter spacetime via the DRG leads 
to a first order equation for the boundary data that exactly corresponds to
the beta function for the coupling in the dual quantum field theory \cite{Nakayama:2013fha}.

%================================
\section{Leading Order Inspiral}
\label{sec:LO}
%================================

The equations of motion in the center-of-mass frame through leading order in the potential (i.e., Newtonian) and radiation reaction forces are~\cite{BurkeThorne:Relativity, Burke:JMathPhys12, Blanchet:LRR}
\begin{align}
	\ba = {} & - \frac{ M}{ r^3 } \br + \frac{ M^2 \nu }{ 15 r^4 } \dot{r} \bigg( \frac{ 136 M}{ r} + 72 \bv^2 \bigg) \br  - \frac{ 8 M^2 \nu }{ 5 r^3 } \bigg( \frac{ 3 M }{ r} + \bv^2 \bigg) \bv   .
\label{eq:eom0PNa}
\end{align}
In terms of polar coordinates, \eqref{eq:eom0PNa} is expressed as
\begin{equation}
\begin{aligned}
\label{LOBT}
	\ddot{r} - r \omega^2 = {} & - \frac{ M }{ r^2 } +  \frac{ 64 M^3 \nu }{ 15 r^4} \dot{r} + \frac{16 M^2 \nu }{ 5 r^3 } \dot{r}^3 + \frac{16 M^2 \nu }{ 5 r } \dot{r} \omega^2  \\
	r \dot{\omega} + 2 \dot{r} \omega = {} &  - \frac{ 24 M^3 \nu }{ 5 r^3 } \omega - \frac{ 8 M^2 \nu }{ 5 r^2 } \dot{r}^2 \omega - \frac{ 8 M^2 \nu }{ 5 } \omega^3 
\end{aligned}
\end{equation}
where $\omega(t) = \dot{\phi}(t)$ is the binary's orbital angular frequency.
The orbital plane does not precess and the motion is described fully by the binary's separation $r(t)$ and the orbital phase $\phi(t)$.

We  will solve these equations perturbatively in the PN expansion.
Of course, we are ignoring the $1$PN and $2$PN conservative forces that should be included for a consistent description through $2.5$PN order. Nevertheless, it is sufficient to use~\eqref{LOBT} for our purpose of demonstrating the DRG method.

%------------------------------------------------------
\subsection{Perturbations of a background circular orbit}
%------------------------------------------------------

We begin by considering the radiation reaction force to be negligible so that the background 
orbital motion is nearly circular. 
We have chosen these conditions because  it is widely expected that many compact binary sources will have circularized by the time their radiated gravitational waves enter the frequency band of ground-based detectors. 
However, it is straightforward to incorporate eccentricity into the background orbit.

The leading order background circular orbit is described by a constant radius $R_B$ and constant angular frequency $\Omega_B$ with\footnote{The $B$ subscript stands for ``bare'', as opposed to ``renormalized''  $R$, which will be discussed further below.
}
\begin{align}
	\Omega_B^2 = \frac{ M }{ R_B^3}.
\end{align}
We next calculate the deviations of this background orbit due to the leading order radiation reaction force~\eqref{LOBT} by writing $r(t) = R_B + \delta r(t)$ and $\omega(t) = \Omega_B + \delta \omega(t)$
where the perturbations scale with the relative velocity at the initial time $t_0$ as $\delta r \sim v_B^5 R_B$ and $\delta \omega \sim v_B^5 \Omega_B \sim v_B^6 / R_B$. 
Expanding out \eqref{LOBT} to first order in $\delta r$ and $\delta \omega$ gives
\begin{equation}
\begin{aligned}
\label{LO}
	 \delta \ddot r(t)-3 \Omega_B^2 \delta r (t)- 2 R_B \Omega_B \delta \omega (t)  & = 0  \\
	R_B \delta \dot \omega (t) + 2 \Omega_B \delta \dot r(t)  & = -\frac{32\nu}{5}  R_B^6 \Omega_B^7.
\end{aligned}
\end{equation}
Solving for $\delta \omega$ and substituting back into the $\delta r$ equation in \eqref{LO} gives
\begin{align}
	\label{eq:harmosc1}
	\delta \ddot r(t) + \Omega_B^2 \delta r(t) = - \frac{ 64 \nu }{5 } \Omega_B^8 R_B^6 (t-t_0)  .
\end{align}
This equation is simple to solve using the retarded Green's function 
\beq
\label{greens}
G_{\rm ret}(t-t^\prime)= \theta(t-t') \frac{\sin \Omega_B (t-t^\prime)}{\Omega_B}
\eeq
and results in the following general solution,
\begin{equation}
\begin{aligned}
\label{LOsol1a}
	r(t) = {} &  R_B - \frac{ 64  \nu }{ 5 } \Omega_B^6 R_B^6(t-t_0)  + \frac{64 \nu}{5}  \Omega_B^5 R_B^6  \sin \Omega_B (t-t_0) 
		 + A \sin \big( \Omega_B (t-t_0) + \Phi \big)  \\
	\omega(t) = {} &   \Omega_B  + \frac{ 96  \nu }{ 5  }R_B^5\Omega_B^7 (t-t_0)   - \frac{ 128 \nu }{ 5 }R_B^5 \Omega_B^6  \sin \Omega_B (t-t_0)  - \frac{2 \Omega_B A }{ R_B } \sin \big( \Omega_B (t-t_0) + \Phi \big) . 
\end{aligned}
\end{equation}
The last two terms are solutions to the  homogeneous equation of~\eqref{eq:harmosc1} and come with two initial condition parameters, $A$ and $\theta$.
As such, we will redefine our background solution to be
\begin{equation}
\begin{aligned}
\label{back}
	r(t) & = R_B+ A_B \sin \big( \Omega_B (t-t_0) + \Phi_B \big)  \\
	\omega(t) & = \Omega_B-\frac{2\Omega_B A_B}{R_B} \sin \big( \Omega_B (t-t_0) + \Phi_B \big) 
\end{aligned}
\end{equation}
where $A_B$ is related to a small orbital eccentricity, $e_B \sim v^5$, through
\begin{align}
\label{eq:AtoEcc}
	A_B = e_B R_B .
\end{align}
The perturbations consist of two types of pieces. 
The first are secularly growing in time. Since, at a time
\begin{align}
	t - t_0 \sim \frac{1}{\nu \Omega_B^6 R_B^5} \sim \frac{ 1}{\nu v_B^5  \Omega_B}  ,
\label{eq:breakdown1loop}
\end{align}
the perturbation becomes $\calO(1)$, these terms will need to be resummed
in order to determine the long-time behavior of the system.
The remaining terms will be perturbatively small for all times.

%-----------------------------------------
\subsection{Renormalization}
%-----------------------------------------

The first step in the resummation procedure is renormalization. 
This involves absorbing all of the $t_0$ dependence into the ``bare'' paratemeters, i.e. those constants labelled by a subscript $B$.
We write our bare solution as
\begin{equation}
\begin{aligned}
	r(t) = {} &  R_B - \frac{ 64  \nu }{ 5 }  R_B^6 \Omega_B^6 (t-t_0)  + \frac{64 \nu}{5}   R_B^6 \Omega_B^5 \sin \Omega_B (t-t_0) 
		 + A_B \sin \big( \Omega_B (t-t_0) + \Phi_B \big)  \\
	\omega(t) = {} &   \Omega_B  + \frac{ 96  \nu }{ 5  }R_B^5\Omega_B^7 (t-t_0)   - \frac{ 128 \nu }{ 5 }R_B^5 \Omega_B^6  \sin \Omega_B (t-t_0)  - \frac{2 \Omega_B A_B }{ R_B } \sin \big( \Omega_B (t-t_0) + \Phi_B \big) . 
\end{aligned}
\end{equation}
Where we have promoted the integration constants $A$ and $\Phi$ to the status of bare parameters.
Notice that $A_B \sim v^5 R_B$ and implies that $\dot r(t_0) \sim v^5 R_B \Omega_B$. 

Furthermore, we may drop the non-secularly growing sinusoidal terms (which are solutions to the homogeneous first-order equations of motion) 
in the solution for $r(t)$ and $\omega(t)$. 
This amounts to  a shift in the initial conditions, which can be accomplished by the following replacement,
\begin{equation}
\begin{aligned}
	A_B \rightarrow {} & A_B- \frac{64\nu}{5}R_B^6\Omega_B^5 \cos(\Phi_B)  \\
	\Phi_B \rightarrow {} & \Phi_B+\frac{64}{5} \frac{\nu R_B^6 \Omega_B^5}{A_B} \sin(\Phi_B).
\end{aligned}
\end{equation}
The bare solution becomes
\begin{equation}
\label{LOsol1a}
\begin{aligned}
		r(t) = {} &  R_B - \frac{ 64  \nu }{ 5 }  R_B^6 \Omega_B^6 (t-t_0)  + A_B \sin \big( \Omega_B (t-t_0) + \Phi_B \big)  \\
	\omega(t) = {} &   \Omega_B  + \frac{ 96  \nu }{ 5  }R_B^5\Omega_B^7 (t-t_0)  - \frac{2 \Omega_B A_B }{ R_B } \sin \big( \Omega_B (t-t_0) + \Phi_B \big) ,
\end{aligned}
\end{equation}
which satisfies the equations of motion. 
For completeness, the orbital phase $\phi(t)$ is computed from $\omega(t)$ via a simple integration,
\begin{align}
	\phi(t) =  \Phi_B + \Omega_B (t-t_0) 
		%+ \frac{ 128 \nu }{ 5 }R_B^5 \Omega_B^5  \cos \Omega_B (t-t_0) 
		+ \frac{ 48  \nu }{ 5  }R_B^5\Omega_B^7 (t-t_0)^2 + \frac{ 2 A_B}{R_B}  \cos \big( \Omega_B (t-t_0)+\Phi_B \big)  .
\label{LOsol1d}
\end{align}
Notice  that $\Phi_B$ is {\it not} the initial phase, $\phi(t_0)$. Overall constants can be
dropped since they can be removed by a coordinate change without affecting the
equations of motion.  

The four quantities $R_B, \Omega_B, A_B$ and $\Phi_B$ are parameters fixed by the initial data of the problem. However, the initial time $t_0$ is completely arbitrary and we could
have performed the perturbative expansion at a slightly later time, $t'_0 = t_0 + \delta t$, for instance. 
The formal expression of the perturbative solution would have the same form as in \eqref{back} except with a new set of intitial conditions $R'_B, \Omega'_B, A'_B$, and $\Phi'_B$ and with $t_0$ replaced by $t'_0$. 
If $\delta t$ is small then it is straightforward to see  that the initial conditions at $t_0$ are related to those at $t'_0$. 
This time shift can be compensated for  by redefining the bare parameters as
\begin{equation}
\begin{aligned}
\label{eq:RGmotivation1}
	R'_B = {} & R_B - \frac{64\nu}{5} R_B^6 \Omega_B^6  \delta t + \calO(\delta t^2) \\
	\Omega'_B = {} & \Omega_B + \frac{96 \nu}{5} R_B^5 \Omega_B^7 \delta t + \calO(\delta t^2) \\
	\Phi'_B = {} & \Phi_B + \Omega_B \delta t +\calO(\delta t^2)  .
\end{aligned}
\end{equation}
Therefore, the perturbative solution at $t'_0$ is related to that at $t_0$ by redefining the initial conditions in such a way as to preserve the form of the perturbative solution in \eqref{back}. 
In this way, one may ``bootstrap'' the perturbative solutions from one time to any other  and thereby generate the long-time inspiral dynamics up to the PN accuracy of the original perturbative solution~\cite{Kunihiro:1995zt}. 
This process of redefining, or {\it renormalizing}, the initial conditions to ensure the form-invariance of the perturbative solution at different times is at the heart of the DRG method~\cite{Chen:1995ena} and, more generally, renormalization group theory. 

We regard $R_B, \Omega_B, A_B$, and $\Phi_B$ as {\it bare parameters} that depend on
the initial time $t_0$, as suggested in \eqref{eq:RGmotivation1}. 
We may think of $t_0$ as the cut-off in the usual Wilsonian sense.
All physical ``renormalized" quantities are independent of $t_0$.
We relate the bare parameters to their renormalized values $R_R, \Omega_R, A_R$, and $\Phi_R$ through the relations
\begin{align}
\label{eq:bareR1}
	R_B(t_0) & = R_R(\tau)+ \delta_R (\tau, t_0) \\
	\Phi_B(t_0) & = \Phi_R(\tau) +  \delta_\Phi (\tau, t_0) \\
	\Omega_B(t_0) & = \Omega_R(\tau) +\delta_\Omega (\tau, t_0) \\
	A_B(t_0) & = A_R(\tau) +  \delta_A (\tau, t_0) ,
\end{align}
where $\delta_R$, $\delta_\Phi$, $\delta_\Omega$,  and $\delta_A$ are quantities called {\it counter-terms} that are to be determined order-by-order in the process of renormalizing the perturbative solutions in \eqref{back} and \eqref{LOsol1d}. 
The new time parameter $\tau$ is the {\it renormalization scale} and is arbitrary. 
The initial time $t_0$ is like a cut-off scale when regularizing the divergences of a field theory.
However, the perturbative solutions are independent of $\tau$ at any given order in perturbation theory. 

In terms of the renormalized initial parameters, the one-loop result becomes
\begin{align}
\label{LOsol2a}
	r(t) = {} & R_R +\delta_R - \frac{ 64  \nu }{ 5 } R_R^6 \Omega_R^6  (t-t_0) 
		 + (A_R+\delta_A)  \sin \big( (t-t_0) \Omega_R + \Phi_R +\delta_\Phi \big)   \\
\label{LOsol2b}
	\omega(t) = {} & \Omega_R + \delta_\Omega   + \frac{ 96  \nu }{ 5  }R_R^5\Omega_R^7 (t-t_0) 
		  - \frac{ 2 \Omega_R (A_R+\delta_A) }{ R_R }  \sin \big( (t-t_0) \Omega_R + \Phi_R + \delta_\Phi \big)  \\
\label{LOsol2c}
	\phi(t) = {} & \Phi_R + \delta_\Phi + (t-t_0)( \Omega_R + \delta_\Omega)  + \frac{ 48  \nu }{ 5  }R_R^5 \Omega_R^7 (t-t_0)^2  
	  + \frac{ 2 (A_R+\delta_A) }{ R_R } \cos \big((t-t_0)\Omega_R + \Phi_R  + \delta_\Phi \big)
\end{align}
where we have dropped terms of order $v^{10}$.

We introduce the renormalization scale into the above solutions through $t-t_0 = (t-\tau) + (\tau-t_0)$ 
so that~\eqref{LOsol2a}--\eqref{LOsol2c} become
\begin{align}
\label{LOsol10a}
	r(t) = {} & R_R +\delta_R - \frac{ 64  \nu }{ 5 } R_R^6 \Omega_R^6 (t - \tau) - \frac{ 64  \nu }{ 5 } R_R^6 \Omega_R^6  (\tau-t_0)
	%+ \frac{64 \nu}{5}   R_R^6 \Omega_R^5 \sin \Omega_R (t-t_0) 
 %\nonumber \\ & 
 + A_R  \sin \big( (t - \tau) \Omega_R + (\tau-t_0) \Omega_R + \Phi_R +\delta_\Phi \big)   \\
\label{LOsol10b}
	\omega(t) = {} & \Omega_R + \delta_\Omega   + \frac{ 96  \nu }{ 5  }R_R^5\Omega_R^7 (t-\tau) + \frac{ 96  \nu }{ 5  }R_R^5\Omega_R^7 (\tau - t_0)
%	- \frac{ 128 \nu }{ 5 }R_R^5 \Omega_R^6  \sin \Omega_R (t-t_0)
%	 \nonumber \\ &
	   - \frac{ 2 \Omega_R A_R }{ R_R }  \sin \big( (t-\tau) \Omega_R 
+ (\tau - t_0) \Omega_R+ \Phi_R + \delta_\Phi \big)  \\
\label{LOsol10c}
	\phi(t) = {} & \Phi_R + \delta_\Phi +(t-\tau) \Omega_R + (\tau-t_0) \Omega_R + (t - \tau) \delta_\Omega + (\tau - t_0) \delta_\Omega  + \frac{ 48  \nu }{ 5  }R_R^5 \Omega_R^7 (t-\tau)^2 \nonumber \\
		& + \frac{ 96  \nu }{ 5  }R_R^5 \Omega_R^7 (t-\tau) (\tau-t_0) + \frac{ 48  \nu }{ 5  }R_R^5 \Omega_R^7 (\tau - t_0)^2 + \frac{ 2 A_R }{ R_R } \cos \big( (t-\tau)\Omega_R + (\tau -t_0) \Omega_R+ \Phi_R  + \delta_\Phi \big)  .
\end{align}
Renormalization proceeds by fixing the counter-terms at this order in $\varepsilon$ to cancel the pieces that are proportional to powers of $(\tau -t_0)$. 
For instance, inspection of~\eqref{LOsol10a} shows that the counter-term $\delta_R$ is fixed at one-loop order to be\footnote{One is, of course, free to add finite contributions but this just amounts to a shift in $\tau$.} 
\begin{align}
\label{1loopCT}
	\delta_R^{v^5} (\tau, t_0) = {} & \frac{ 64 \nu}{ 5} R_R^6 \Omega_R^6 (\tau -t_0)
\end{align}
where we have written $\delta_R = \delta_R^{v^5} + \calO(\varepsilon^2)$. 
The $\calO(\varepsilon^2)$ term is a two-loop contribution that will be calculated in the next section.
Likewise, the counter term $\delta_\Omega$ is found from~\eqref{LOsol10b} to cancel the term proportional to $\tau-t_0$ so that
\begin{align}
\label{1loopCTb}
	\delta_\Omega^{v^5} (\tau, t_0) = {} & - \frac{ 96  \nu }{ 5  } R_R^5 \Omega_R^7(\tau-t_0)
\end{align}
where we have again written $\delta_\Omega = \delta_\Omega^{v^5} + \calO(\varepsilon^2)$.

Then, substituting these counter-terms into~\eqref{LOsol10c}, we find that the perturbative solution for $\phi(t)$  becomes
\begin{align}
	\phi(t)  = {} & \Phi_R + \delta_\Phi + (t-\tau) \Omega_R + (\tau -t_0) \Omega_R + \frac{ 48  \nu }{ 5  }R_R^5 \Omega_R^7 (t-\tau)^2 - \frac{ 48  \nu }{ 5  }R_R^5 \Omega_R^7 (\tau-t_0)^2  \nonumber \\
		 & + \frac{ 2 A_R }{ R_R } \cos \big( (t-\tau)\Omega_R + (\tau -t_0) \Omega_R + \Phi_R  + \delta_\Phi \big) 
\end{align}
Notice that the term proportional to $(t-\tau)(\tau-t_0)$ automatically cancels out of the equation, which turns out to be an important check of self-consistency as we shall see in the next section. We see that we will need an additional counter-term for $\phi_B$.
Choosing
\begin{equation}
	\delta_\Phi (\tau, t_0) =  -\Omega_R(\tau-t_0) + \frac{ 48  \nu }{ 5  } R_R^5  \Omega_R^7 (\tau-t_0)^2  + \calO(\varepsilon^2 )
\label{1loopCTc}
\end{equation}
removes the last remaining secular terms, even those appearing inside the oscillating terms in~\eqref{LOsol10a} and~\eqref{LOsol10b} to this order in $\varepsilon$.
We then are left with the renormalized perturbative solutions 
\begin{align}
\label{LOsol3a}
	r(t)  = {} & R_R  - \frac{ 64  \nu }{ 5 }R_R^6 \Omega_R^6  (t-\tau) + A_R  \sin \big(  (t-\tau) \Omega_R + \Phi_R \big)   \\
\label{LOsol3b}
	\omega(t) = {} & \Omega_R + \frac{ 96  \nu }{ 5  }R_R^5\Omega_R^7 (t-\tau)  
		 - \frac{ 2 \Omega_R A_R }{ R_R} \sin \big( (t-\tau) \Omega_R + \Phi_R \big)  \\
\label{LOsol3c}
	\phi(t) = {} & \Phi_R+\Omega_R(t-\tau)  + \frac{ 48  \nu }{ 5  }R_R^5\Omega_R^7(t-\tau)^2  
		+ \frac{ 2 A_R }{ R_R } \cos \big( (t-\tau)\Omega_R + \Phi_R \big).
\end{align}
Since $\tau$ is arbitrary we will choose it to equal $t$ when we consider the physical solution so as to minimize all of the secular terms giving
\begin{align}
\label{LOsol4a}
	r(t)  = {} & R_R(t)  + A_R(t)  \sin  \Phi_R(t)  \\
\label{LOsol4b}
	\omega(t) = {} & \Omega_R(t) - \frac{ 2 \Omega_R(t) A_R(t) }{ R_R(t) } \sin \Phi_R(t)  \\
\label{LOsol4c}
	\phi(t) = {} & \Phi_R(t) + \frac{ 2 A_R(t) }{ R_R(t) } \cos  \Phi_R (t).
\end{align}
This step is akin to scale-setting in the context of canonical RG flows.

%-----------------------------------------
\subsection{The Renormalization Group solution}
%-----------------------------------------

The time dependence of the renormalized initial data is found by noting that the bare parameters are independent of the arbitrary scale $\tau$ so that $d R_B (t_0)  / d\tau = 0$ and likewise for the other three initial parameters. 
For example, recall that the bare parameter $R_B$ is given in~\eqref{eq:bareR1} by
\begin{align}
	R_B (t_0) = R_R(\tau) + \delta_R (\tau, t_0) = R_R + \frac{ 64\nu}{ 5} R_R^6 \Omega_R^6 (\tau -t_0) + \calO(\varepsilon^2)
\end{align}
so that
\begin{align}
\label{eq:RGeqnTest1}
	0 = \frac{ d R_B(t_0) }{ d\tau } = \frac{ d R_R (\tau) }{ d\tau } + \frac{ 64\nu}{5} R_R^6 \Omega_R^6 + \frac{384 \nu }{ 5} R_R^5 \Omega_R^6 (\tau-t_0) \frac{ dR_R(\tau) }{ d\tau} + \frac{ 384\nu}{ 5} R_R^6 \Omega_R^5 (\tau - t_0) \frac{ d\Omega_R(\tau) }{ d\tau} + \calO(v_R^{10})  .
\end{align}
It is easy to see that solving perturbatively for $dR_R / d\tau$ leaves us with
\begin{align}
\label{eq:RGeqns1a}
	\frac{d}{d\tau} R_R(\tau) = {} & - \frac{64 \nu}{5}  R^6_R(\tau) \Omega^6_R(\tau)+....
\end{align}
since the last two terms in~\eqref{eq:RGeqnTest1} are higher order corrections.
Repeating these steps for the remaining initial data yields a total of four {\it renormalization group equations} describing the RG {\it flow}, or {\it trajectory}, of the initial conditions
\begin{align}
\label{eq:RGeqns1b}
	\frac{d}{d\tau}\Omega_R(\tau) = {} & \frac{96 \nu}{5}  R_R^5(\tau) \Omega^7_R(\tau) , \\
\label{eq:RGeqns1c}
	\frac{d}{d\tau} \Phi_R (\tau) = {} & \Omega_R (\tau), \\
\label{eq:RGeqns1d}
	\frac{d}{d\tau} A_R (\tau) = {} & 0  .
\end{align}
The right sides of these equations are called beta ($\beta$) functions in field theory.
The solutions to the RG equations~\eqref{eq:RGeqns1a}-\eqref{eq:RGeqns1d} are easily found by integrating from $\tau = t_i$ to $\tau = t$,
\begin{align}
\label{LOa}
	R_R(t) = {} & \bigg( R_R^4(t_i) - \frac{256 \nu}{5} M^3 (t-t_i) \bigg)^{1/4} \\ 
\label{LOb}
	\Omega_R(t) = {} & \Omega_R(t_i) \left( \frac{R_R(t_i)}{R_R(t)} \right)^{3/2}  \\
\label{LOc}
	\Phi_R(t) = {} & \Phi_R(t_i) + \frac{R^{5/2}_R(t_i) - R_R^{5/2}(t) }{ 32 \nu M^{5/2}} \\
\label{LOd}
	A_R(t) = {}& A_R (t_i).
\end{align}
These are nothing but the textbook orbit-averaged solutions (see for instance \cite{Maggiore}). 
Thus, the difference between the DRG solutions and the orbit-averaged solutions are
the sinusoidal terms in \eqref{LOsol4a}-\eqref{LOsol4c}. Note that these terms do not have constant periods and thus orbit averaging will not set them strictly to zero. The lack of a definite period is another
weakness of the averaging procedure \cite{pound2005limitations,pound2008osculating}

As can be seen from \eqref{LOb}, the quantity $R_R^3(t) \Omega_R^2(t)$ is an invariant along the RG trajectory. 
This constant is just equal to $M$. 
Other RG invariants can be found from these relations that are not so trivial, including
\begin{align}
	R^4_R(t) + \frac{256 \nu}{5} M^3 t & = {\rm constant} \\
	\Phi_R(t) + \frac{ R_R^{5/2}(t) }{ 32 \nu M^{5/2} } & = {\rm constant} .
\end{align}
The expressions in \eqref{LOa}-\eqref{LOd}, combined with the renormalized solutions in~\eqref{LOsol4a}-\eqref{LOsol4c}, give the resummed solution to the $0$PN inspiral dynamics valid up to times $t-t_i$ of
order $1 / (\nu v_R^5(t) \Omega_R(t) )$. Note that the initial radial velocity depends on $A_R(t_i)$ and $\Phi_R(t_i)$ at this order via the relation
\begin{align}
\label{eq:rdot1a}
	\dot r(t_i) = A_R(t_i) \Omega_R(t_i) \cos \Phi_R(t_i) - \frac{64 \nu}{5} R_R(t_i)^6 \Omega_R(t_i)^6. 
\end{align}

For the purposes of comparison, we next find the numerical solution of an equal mass compact binary inspiral where the total mass is $M=1$.
Specifically, we choose the following initial data at $t_i = 0$ for demonstration purposes,
\begin{align}
\label{eq:initdata}
	\begin{array}{r c l}
		\phi(0) & = & 0 \\
		\omega(0) & = & 10^{-2} / M \\
		r(0) & = & ( M / \omega(0)^2 )^{1/3} = 10^{4/3} M \\
		\dot{r}(0) & = & 0 .
	\end{array}
\end{align}
%which we impose for all our solutions, both numerical and analytical.
Notice that the typical speed scale is $v \sim r(0) \omega(0) \approx 0.2$ and $v^5 \sim 5 \times 10^{-4}$, which are manageable numbers for numerical studies and is why we have chosen them.
To relate these initial conditions to the parameters $R_R(t_i)$, $\Omega_R(t_i)$, $\Phi_R(t_i)$, and $A_R(t_i)$, we set \eqref{LOsol3a}-\eqref{LOsol3c} and the time derivative of~\eqref{LOsol3a} at $t_i=0$ equal to the above initial data. 
This yields four equations in the four parameters, which we solve numerically.
Recall that, $A_R(0) = e_R(0) R_R(0)$ is proportional to the initial eccentricity $e_R(0)$, which we took to be $\calO(v^5)$.

%%%%%%%%%%
% The text below is for a binary neutron star inspiral and is commented out.
%%%%%%%%%%
\begin{comment}

For the purposes of comparison, we next find the numerical solution of an equal mass binary with component masses of $1.4 M_\odot$ (i.e., neutron stars) starting from an initial orbital frequency of $5$Hz corresponding to a $10$Hz gravitational wave.
Specifically, we choose the following initial data at $t_i = 0$,
\begin{align}
\label{eq:initdata}
	\left. 
	\begin{array}{r c l}
		\Omega_R(0) & = & 10\pi {\rm ~rad/s} \\
		R_R (0) & = & \left( \displaystyle\frac{ (1.4 + 1.4) M_\odot }{ \Omega_R(0)^2 } \right)^{1/3} \\
		\Phi_R(0) & = & 0 \\
		A_R(0) & = & \displaystyle\frac{64 \nu}{5} R_R(0)^6 \Omega_R(0)^5
	\end{array}
	\right\}  
	~~ \Longrightarrow ~~
	\left\{
	\begin{array}{r c l}
		r(0) & = & R_R(0) \\
		\dot{r}(0) & = & 0 \\
		\omega(0) & = & \Omega_R(0) \\
		\phi(0) & = & \Phi_R(0) + \displaystyle\frac{2 A_R(0)}{R_R(0)}
	\end{array}
	\right.
\end{align}
which we impose for all our solutions, both numerical and analytical.
The initial condition $A_R(0)$ comes from imposing $\dot{r}(0) = 0$ in \eqref{eq:rdot1a} and solving for $A_R(0)$. 
Of course, $A_R(0) = e_R(0) R_R(0)$ is proportional to the initial eccentricity $e_R(0)$, which we recall is taken to be $\calO(v_R^5)$.

\end{comment}
%%%%%%%%%%
%%%%%%%%%%

\begin{figure}
	\includegraphics[width=0.48\columnwidth]{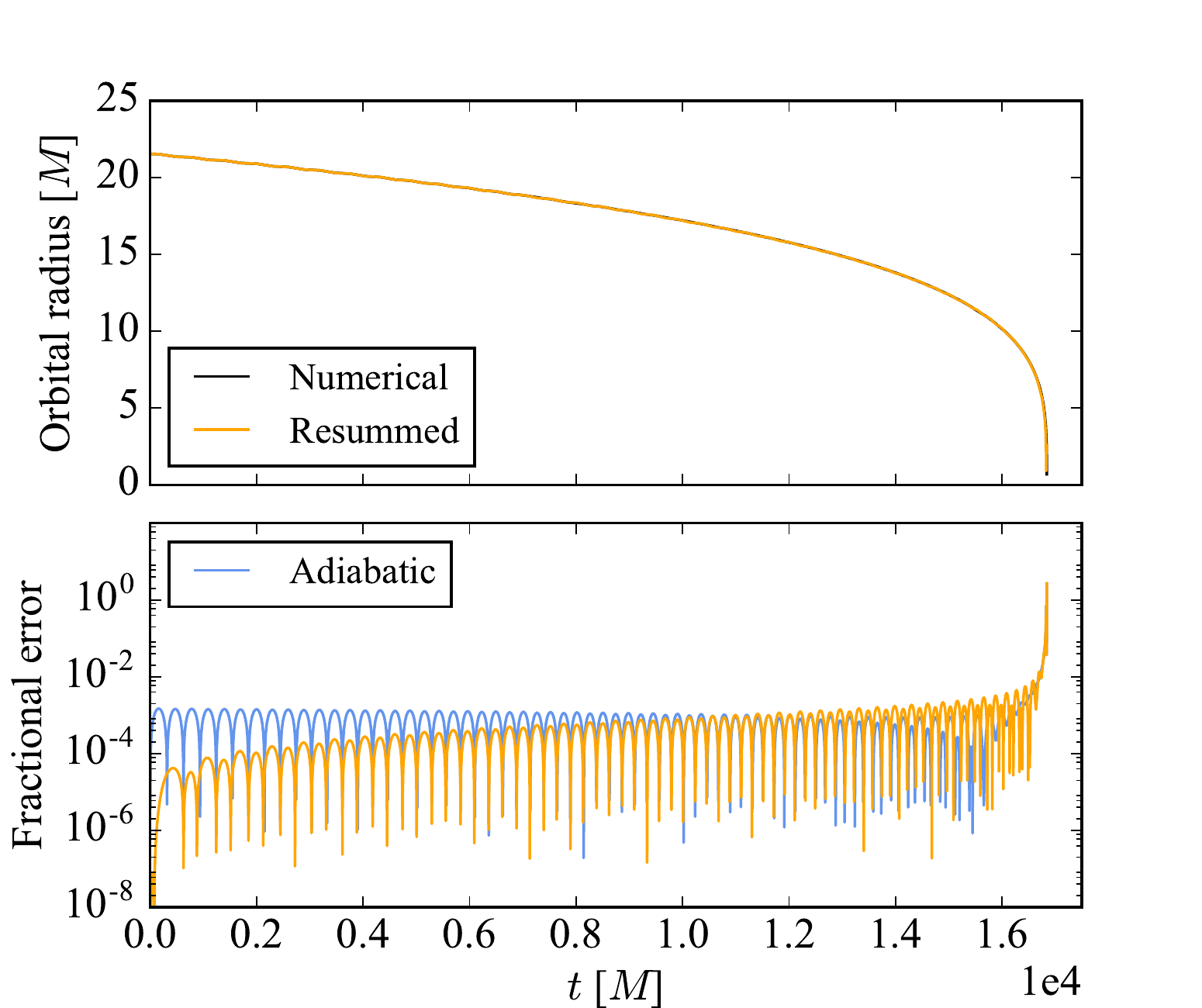}
	\includegraphics[width=0.48\columnwidth]{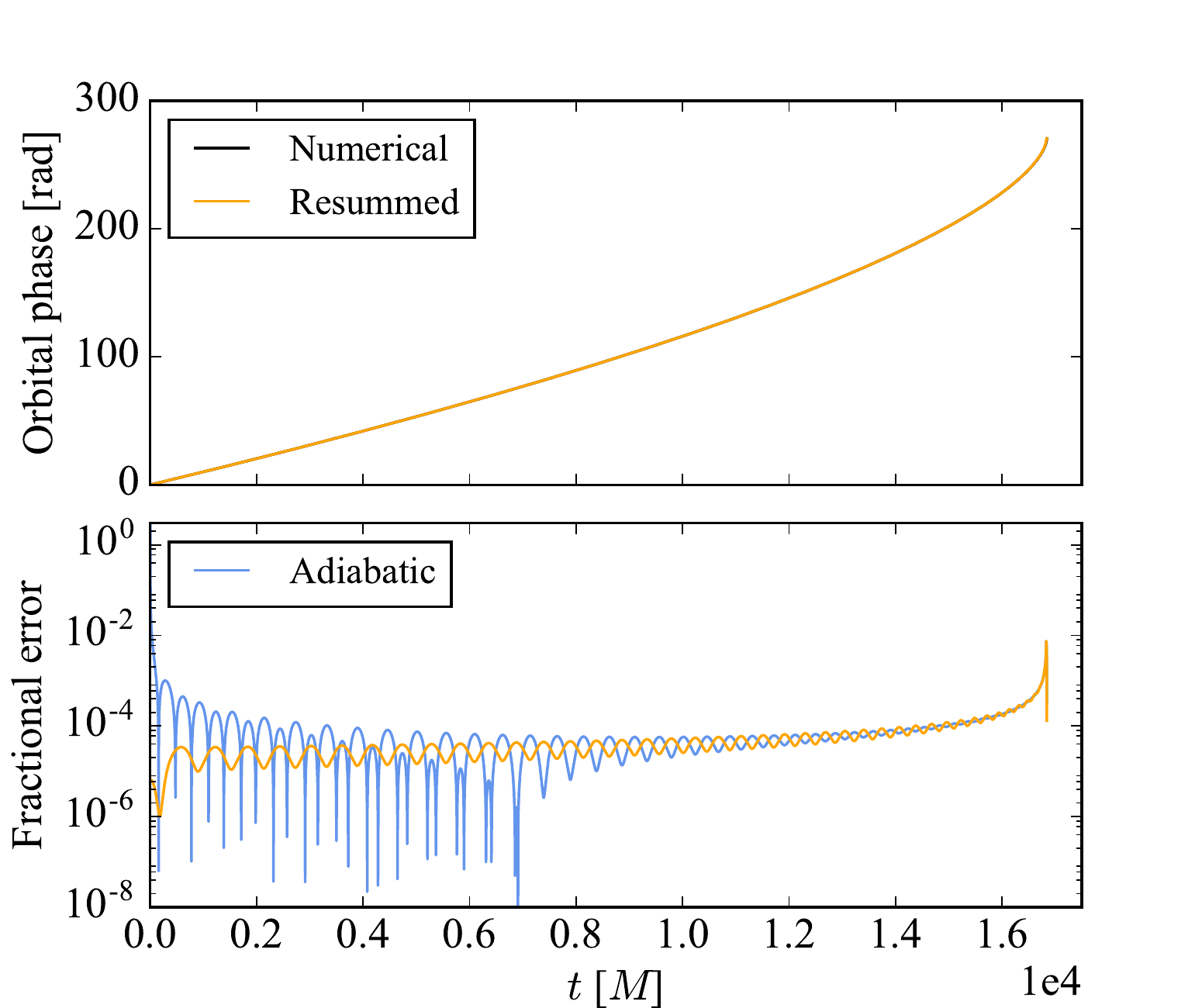}
	\label{fig:comparison1}
	\caption{{\bf Top panels}: Orbital radius and phase versus time for an equal-mass compact binary inspiral
		with initial data given in \eqref{eq:initdata}.
		The numerical solution of the $0$PN equations of motion (black) 
		and the renormalization group
		resummed solution (orange) are shown.
		{\bf Bottom panels}: Fractional errors for the orbital radius and phase, respectively, between the numerical solution and the resummed (orange) and adiabatic or orbit-averaged (blue) approximate solutions.} 
\end{figure}

In Fig.~\ref{fig:comparison1} we compare the numerical solution (black) to our RG resummed solution (orange).
The top left (right) panel shows these solutions for the orbital radius (phase). 
%The resummed solution seems to agree with the numerical solution very well over the entire inspiral. 
The bottom panels show the fractional errors for the orbital radius and phase solutions, respectively. 
The orange (blue) curves show the fractional errors between the RG resummed (adiabatic, orbit-averaged) and numerical solutions. 

The adiabatic solutions come from solving the flux-balance equations, which are averaged over the orbital period~\cite{Blanchet:LRR}. 
It should be noted that the adiabatic solutions contain an ambiguity in specifying the initial data because of the orbit-averaging procedure, as discussed in~\cite{pound2008osculating, pound2005limitations}, which can be seen at early times in the bottom, right panel of Fig.~\ref{fig:comparison1}.
In addition, for orbits with larger eccentricities it is not clear which oscillations the adiabatic approximation should remove (e.g., those parameterized by coordinate time, eccentric anomaly, true anomaly, or mean anomaly), which becomes important for periastron advance when PN corrections are included.
As such, comparisons to the adiabatic approximation should be regarded as more qualitative rather than quantitative, perhaps. 
With these comments in mind, we remark that the DRG method provides a systematic procedure for deriving unambiguous predictions for the compact binary's real-time evolution.

In Sec.~\ref{sec:twoloops} we will improve the accuracy of the resummed perturbative solution to~\eqref{LOBT}, especially at late times, by including second order corrections in $\varepsilon$, which will induce an RG flow for the renormalized oscillation amplitude $A_R$ (i.e., the orbital eccentricity).

%------------------------------------
\subsection{Estimating errors of the resummed solutions}
\label{sec:errors}
%------------------------------------

%The RG resummed perturbative solutions are approximations that are valid for much longer times than solutions from naive perturbation theory. 
%Here, we estimate the accuracy of the resummed solutions following~\cite{Chen:1995ena}.

The bare perturbative solutions in~\eqref{LOsol1a}-\eqref{LOsol1d} are accurate up to $\calO(v_B^{10})$
corrections when ignoring higher order PN corrections that we did not originally include in the equations of motion in \eqref{LOBT}. 
When renormalizing the integration constants the error being made in the perturbation theory is $\calO (v_R^{10})$ because all bare parameters are written in terms of their renormalized values plus higher order counter-terms.

Next, we recall that $r(t)$, $\omega(t)$, and $\phi(t)$ are independent of $\tau$ so that differentiating the radial solution, for example, implies that
\begin{align}
	0 = \frac{ dR_R}{d\tau } + \frac{ 64}{5} \nu R_R^6 \Omega_R^6 + A_R \bigg( \frac{ d\Phi_R }{d\tau} - \Omega_R \bigg) \cos \big( \Omega_R (t - \tau) + \Phi_R \big) + \calO( v_R^{10} R_R \Omega_R)
\end{align}
where we have included the error term.
The extra factor of $R_R$ in the error term is to ensure the correct dimensions and scaling for the radial solution while the factor of $\Omega_R$ is the reciprocal of the orbital time scale from the $\tau$ derivative.
Of course, the RG equations in \eqref{eq:RGeqns1a} and \eqref{eq:RGeqns1c} tell us that this is satisfied identically but the error term implies that the RG equations should be written more completely as
\begin{align}
	\frac{ dR_R}{d\tau} = - \frac{ 64}{5} \nu R_R^6 \Omega_R ^6 + \calO( v_R^{10} R_R \Omega_R)
\end{align}
and similarly for the other ones. Therefore, the RG solutions are determined up to $\calO (v_R^{10} \Omega_R (t-t_i) )$ corrections and the resummed perturbative solutions are valid until times
\begin{align}
	t - t_i \sim \frac{ 1}{ v_R^{10} \Omega_R } \sim \frac{ 1 }{ R_R^{10} \Omega_R^{11}}  .
\end{align} 
Notice that this elapsed time of validity is measured with respect to the renormalized integration constants at the initial time $t_i$.

%================================
\section{Going to ``Two Loops":  Two insertions of radiation reaction}
\label{sec:twoloops}
%================================

We now show how to include two insertions of the leading order radiation reaction force. In so doing we will show how to renormalize to $\calO(\varepsilon^2)$ in the DRG formalism. 
Following the field theory terminology we call this a ``two-loop" calculation, despite the fact that all our calculations are more akin to ``tree" level Feynman diagrams.
Indeed, the DRG calculations can be couched in terms of these diagrams by thinking of the background (circular) orbit as a source insertion and treating $r(t)$ and $\omega(t)$ as two distinct one-dimensional fields. 
However, it is not clear that Feynman diagrams are of much utility for us, though they may help keep track of the systematics as one goes to higher orders.

We work in what is known as ``bare perturbation theory.'' In this way of organizing the calculation\footnote{At leading order the distinction between ``bare" and "renormalized" perturbation theory is nominal.}
we work with only bare parameters at arbitrary order and then fix the counter-terms {\it a posteriori}, as we did in the previous section.
We will see that at second order there will be a non-trivial set of consistency checks of the
calculation.

We begin by introducing the second order notation
\begin{align}
	\label{notation}
	r = {} & R_B+\delta r(t) + \delta \kappa (t) \nn \\
	\omega = {} & \Omega_B + \delta \omega(t) + \delta \rho(t)
\end{align}
where $\delta \kappa \sim v_B^{10}R_B$ and $\delta \rho \sim v_B^{10} \Omega_B \sim v_B^{11}/R_B$ and the first order solutions were calculated in the previous section.
The equations of motion for $\delta \kappa$ and $\delta \rho$ are
\begin{align}
\label{NLO}
	\delta \ddot \kappa (t)-3\Omega_B ^2 \delta \kappa (t)  = {} & \frac{112}{15} \nu  R_B^5 \Omega_B ^6 \delta \dot r(t)-\frac{3 \Omega_B ^2 }{R_B}  \delta \dot r^2(t)
		 +R_B \delta \omega^2 (t) + 2 R_B \Omega_B  \delta \rho (t) 
		 + 2 \Omega_B \delta r(t) \delta \omega (t)  \\
	\delta \dot \rho (t) + \frac{ 2 \Omega_B }{R_B}  \delta \dot\kappa (t) = {} & -\frac{48}{5} \nu  R_B^5 \Omega_B ^6 \delta \omega (t) + 8 \nu  R_B^4 \Omega_B ^7 \delta r(t) 
		 - \frac{2}{R_B} \delta \omega (t) \delta \dot r(t) - \frac{2}{R_B} \delta r(t) \delta \dot \omega (t) 
		 - \frac{2 \Omega_B}{R_B^2}  \delta r(t)  \delta \dot r(t).
\end{align}
The solution for the second order radial perturbation $\delta \kappa$ is given by 
\begin{align}
\label{d2}
	\delta \kappa(t) = {} & - \frac{3}{2} \frac{A_B^2 }{ R_B } + \frac{ 29\,696}{75} \nu^2 R_B^{11} \Omega_B^{10} - \frac{ 6144}{25} \nu^2 R_B^{11} \Omega_B^{12} (t-t_0)^2 + \frac{ 272}{5} \nu A_B R_B^5 \Omega_B^5 \cos \Phi_B + \frac{ 3 A_B^2 }{ R_B } \cos 2 \Phi_B \nonumber \\
		& + \frac{3}{2} \frac{A_B^2}{R_B} \cos \Omega_B(t-t_0) - \frac{ 29\,696}{75} \nu^2 R_B^{11} \Omega_B^{10} \cos \Omega_B (t-t_0) - \frac{32}{3} \nu A_B R_B^5 \Omega_B^5 \cos \big( \Phi_B - \Omega_B(t-t_0) \big) \nonumber \\
		& - \frac{5}{4} \frac{ A_B^2 }{R_B} \cos \big( 2\Phi_B - \Omega_B (t-t_0)  \big) - \frac{656}{15} \nu A_B R_B^5 \Omega_B^5 \cos \big( \Phi_B + \Omega_B (t-t_0) \big) \nonumber \\
		& + \frac{48}{5} \nu A_B R_B^5 \Omega_B^7 (t-t_0)^2 \cos \big( \Phi_B + \Omega_B(t-t_0) \big) + \frac{1}{2} \frac{ A_B^2 }{ R_B} \cos \big( 2 \Phi_B + 2 \Omega_ B(t-t_0) \big) \nonumber \\
		& - \frac{ 9}{4} \frac{ A_B^2 }{R_B} \cos \big( 2\Phi_B + \Omega_B (t-t_0) \big) - \frac{ 496}{15} \nu A_B R_B^5 \Omega_B^6 (t-t_0) \sin \big( \Phi_B + \Omega_B (t-t_0) \big)
\end{align}
while that for $\delta \rho$ is
\begin{align}
	\delta \rho(t) = {} & \frac{ 3 A_B^2 \Omega_B }{ R_B } - \frac{ 59\,392}{ 75} \nu^2 R_B^{10} \Omega_B^{11} + \frac{ 16\,896}{ 25} \nu^2 R_B^{10} \Omega_B^{13} (t-t_0)^2 - \frac{ 408}{5} \nu A_B R_B^4 \Omega_B^6 \cos \Phi_B - \frac{9}{2} \frac{ A_B^2 \Omega_B }{R_B^2} \cos 2 \Phi_B \nonumber \\
		& - \frac{ 3 A_B^2 \Omega_B }{ R_B^2 } \cos \Omega_B (t-t_0) + \frac{ 59\,392}{75} \nu^2 R_B^{10} \Omega_B^{11} \cos \Omega_B (t-t_0) + \frac{ 64}{3} \nu A_B R_B^4 \Omega_B^6 \cos \big( \Phi_B - \Omega_B (t-t_0) \big) \nonumber \\
		& + \frac{5}{2} \frac{ A_B^2 \Omega_B }{R_B^2} \cos \big( 2\Phi_B - \Omega_B(t-t_0) \big) + \frac{ 904}{15} \nu A_B R_B^4 \Omega_B^6 \cos \big( \Phi_B + \Omega_B (t-t_0) \big) \nonumber \\
		& - \frac{ 96}{5} \nu A_B R_B^4 \Omega_B^8 (t-t_0)^2 \cos \big( \Phi_B + \Omega_B (t-t_0) \big) - \frac{5}{2} \frac{ A_B^2 \Omega_B }{ R_B^2 } \cos \big( 2\Phi_B + 2 \Omega_B (t-t_0) \big) \nonumber \\
		& + \frac{ 9}{2} \frac{ A_B^2 \Omega_B }{ R_B^2 } \cos \big( 2 \Phi_B + 2 \Omega_B (t-t_0)\big) + \frac{ 32}{15} \nu A_B R_B^4 \Omega_B^7 (t-t_0) \sin \big( \Phi_B + \Omega_B (t-t_0) \big)
\end{align}
As in the previous section for the one-loop calculation, we can shift the initial parameters so as to remove the redundant pieces that are finite (i.e., non-secular) homogenous solutions. 
It is straightforward to show with some algebra and trigonometric identities that the following shift 
\begin{equation}
\begin{aligned}
\label{eq:twoLoopShifts1}
	A_B \to {} & A_B - \frac{ 15}{4} \frac{ A_B^2 }{ R_B } \sin \Phi_B + \frac{ 29\,696}{75 } \nu^2 R_B^{11} \Omega_B^{10} \sin \Phi_B + \frac{ 32}{3} \nu A_B R_B^5 \Omega_B^5 \sin 2 \Phi_B + \frac{ 5}{4} \frac{ A_B^2 }{ R_B } \sin 3 \Phi_B  \\
	\Phi_B \to {} & \Phi_B + \frac{3}{4} \frac{ A_B}{R_B} \cos \Phi_B + \frac{ 29\,696}{75} \frac{ \nu^2 R_B^{11} \Omega_B^{10} }{ A_B } \cos \Phi_B + \frac{32}{3} \nu R_B^5 \Omega_B^5 \cos 2 \Phi_B + \frac{ 5 }{ 4} \frac{ A_B }{R_B} \cos 3 \Phi_B \\
	R_B \to {} & R_B + \frac{ 2 A_B^2 }{R_B} - \frac{ 118\,784}{225} \nu^2 R_B^{11} \Omega_B^{10} - \frac{272}{5} \nu A_B R_B^5 \Omega_B^5 \cos \Phi_B - \frac{ 3 A_B^2 }{R_B} \cos 2 \Phi_B \\
	\Omega_B \to {} & \Omega_B - \frac{ 3 A_B^2 \Omega_B }{ R_B } + \frac{ 59\,392}{75} \nu^2 R_B^{10} \Omega_B^{11} + \frac{ 408}{5} \nu A_B R_B^4 \Omega_B^6 \sin \Phi_B - \frac{ 9}{2} \frac{ A^2_B \Omega_B }{R_B} \cos 2 \Phi_B 
\end{aligned}
\end{equation}
removes the redundant, finite terms\footnote{These shifts have some freedom parametrized by a constant $\mu$ that should be fixed. For the shifts in~\eqref{eq:twoLoopShifts1}, we have chosen a scheme so as to keep the resulting two-loop RG equations as simple as possible, which is equivalent to choosing $\mu$ so as to remove all of the finite, $t$-independent pieces in $\delta \rho(t)$. Of course, one is free to choose other values for $\mu$, which changes the ensuing RG equations and perturbative expressions but in a way that doesn't change the predictions for the physical quantities, $r(t)$ and $\phi(t)$.}.
We are then left with the following expressions for the perturbative solutions at $\calO(v^{10})$
\begin{align}
	\delta \kappa (t) = {} & \frac{1}{2} \frac{ A_B^2 }{R_B} - \frac{ 29\,696}{75} \nu^2 R_B^{11} \Omega_B^{10} - \frac{6144}{25} \nu^2 R_B^{11} \Omega_B^{12} (t-t_0)^2 - \frac{656}{15} \nu A_B R_B^5 \Omega_B^5 \cos \big( \Phi_B + \Omega_B (t-t_0) \big)  \nonumber \\ 
		& + \frac{ 48}{5} \nu A_B R_B^5 \Omega_B^7 (t-t_0)^2 \cos \big( \Phi_B + \Omega_B(t-t_0) \big) + \frac{1}{2} \frac{ A_B^2 }{R_B} \cos \big( 2 \Phi_B + 2 \Omega_B (t-t_0) \big)  \nonumber \\
		& - \frac{ 496}{15} \nu A_B R_B^5 \Omega_B^6 (t-t_0) \sin \big( \Phi_B + \Omega_B (t-t_0) \big)  \\
	\delta \rho(t) = {} &  \frac{ 16\,896}{25} \nu^2 R_B^{10} \Omega_B^{13} (t-t_0)^2 + \frac{ 904}{15} \nu A_B R_B^4 \Omega_B^6 \cos \big( \Phi_B + \Omega_B (t-t_0) \big) \nonumber \\
		& - \frac{ 96}{5} \nu A_B R_B^4 \Omega_B^8 (t-t_0)^2 \cos \big( \Phi_B + \Omega_B (t-t_0) \big) - \frac{5}{2} \frac{ A_B^2 \Omega_ B}{R_B^2} \cos \big( 2 \Phi_B + 2 \Omega_B (t-t_0) \big) \nonumber \\
		& + \frac{32}{15} \nu A_B R_B^4 \Omega_B^7 (t-t_0) \sin \big( \Phi_B + \Omega_B (t-t_0) \big)
\end{align}
which are easily shown to  satisfy the equations of motion to the order we a working.

%-------------------------------------------------------
\subsection{Renormalization}
\label{sec:twoLoopRenormalization}
%-------------------------------------------------------

Starting from the bare perturbative solutions for $\delta \kappa (t)$ and $\delta \rho(t)$ we next renormalize the initial parameters of the system to absorb the secular divergences as we did above at 1-loop.

Additional contributions enter at $\calO(v^{10})$ that come from expanding out the bare parameters of the one-loop contribution to $r(t)$ as well as the $\delta^{v^{10}}_R$ counter term that comes from the background piece, $R_B$. 
The totality of those pieces together with the expression for $\delta \kappa(t)$ gives the full renormalized $v^{10}$ contribution to $r(t)$, which we call $r_{v^{10}}(t)$.
Using the expressions for the one-loop counter-terms given in \eqref{1loopCT}, \eqref{1loopCTb}, and \eqref{1loopCTc} and introducing the renormalization scale $\tau$ through $t-t_0 = (t - \tau) + (\tau -t_0)$, we find that the full $\calO(v^{10})$ contribution to the perturbative radial solution is
\begin{align}
	r_{v^{10}}(t) = {} & \frac{1}{2} \frac{ A_R^2 }{R_R} - \frac{ 29\,696}{75} \nu^2 R_R^{11} \Omega_R^{10} - \frac{6144}{25} \nu^2 R_R^{11} \Omega_R^{12} \big[ (t-\tau)^2  - (\tau - t_0) ^2 \big]  - \frac{656}{15} \nu A_R R_R^5 \Omega_R^5 \cos \big( \Phi_R + \Omega_R (t-\tau) \big)  \nonumber \\ 
		& + \frac{ 48}{5} \nu A_R R_R^5 \Omega_R^7 (t-\tau)^2 \cos \big( \Phi_R + \Omega_R (t-\tau) \big) + \frac{1}{2} \frac{ A_R^2 }{R_R} \cos \big( 2 \Phi_R + 2 \Omega_R (t-\tau) \big) \nonumber \\
		&  - \frac{ 496}{15} \nu A_R R_R^5 \Omega_R^6 \big[ (t-\tau) + (\tau - t_0) \big] \sin \big( \Phi_R + \Omega_R (t-\tau) \big)  +\delta_R^{v^{10}}  + \delta_A^{v^{10}} \sin \big( \Phi_R+  (t - \tau) \Omega_R \big) 
\label{eq:roft151}
\end{align}
where, as usual, we are ignoring terms that are beyond $v^{10}$.

The  $\calO(v^{10})$ counter-terms for $R$ and $A$ are given by
\begin{equation}
\begin{aligned}
	\delta_R^{v^{10}} & = - \frac{6144}{25}  \nu^2  R_R^{11} \Omega_R ^{12} (\tau-t_0)^2 \\
	\delta_A^{v^{10}} & =   \frac{496}{15}A_R \nu  R_R^5 \Omega_R ^6  (\tau-t_0) .
\end{aligned}
\end{equation}
In calculating~\eqref{eq:roft151} we encounter terms proportional to $(t - \tau) (\tau -t_0)$ between the linear and quadratic terms. 
The fact that such cross-terms cancel when using the expressions for the one-loop counter terms constitutes a consistency check because otherwise there would be residual $t_0$ contributions surviving that would be akin to having a ``non-renormalizable" field theory. 
The renormalized, finite contribution to the second order radial perturbation is then given by 
\begin{align}
\label{dk}
	r_{v^{10}(t)} = {} &  \frac{1}{2} \frac{ A_R^2 }{R_R} - \frac{ 29\,696}{75} \nu^2 R_R^{11} \Omega_R^{10}  - \frac{656}{15} \nu A_R R_R^5 \Omega_R^5 \cos \Phi_R  + \frac{1}{2} \frac{ A_R^2 }{R_R} \cos 2 \Phi_R 
\end{align}
and we have again used the scale-setting, $\tau=t$.

As with the radial solution, additional contributions contribute to $\delta \rho(t)$ at $\calO(v^{10})$ that come from expanding out the bare parameters of the one-loop contribution to $\omega(t)$ as well as the $\delta^{v^{10}}_\Omega$ counter term that comes from the background piece, $\Omega_B$. 
The totality of those pieces together with the expression for $\delta \rho(t)$ gives the full renormalized $v^{10}$ contribution to $\omega(t)$, which we call $\omega_{v^{10}}(t)$.
Using the expressions for the one-loop counter-terms given in \eqref{1loopCT}, \eqref{1loopCTb}, and \eqref{1loopCTc} and introducing the renormalization scale $\tau$ through $t-t_0 = (t - \tau) + (\tau -t_0)$, we find that the full $\calO(v^{10})$ contribution to the perturbative angular frequency solution is
\begin{align}
\label{drho}
	\omega_{v^{10}}(t) = {} & \frac{ 16\,896}{25} \nu^2 R_R^{10} \Omega_R^{13} \big[ (t-\tau)^2 - (\tau - t_0^2) \big]  + \frac{ 904}{15} \nu A_R R_R^4 \Omega_R^6 \cos \big( \Phi_R + \Omega_R (t-\tau) \big) \nonumber \\
		& - \frac{ 96}{5} \nu A_R R_R^4 \Omega_R^8 (t-\tau)^2  \cos \big( \Phi_R + \Omega_R (t-\tau) \big) - \frac{5}{2} \frac{ A_R^2 \Omega_R}{R_R^2} \cos \big( 2 \Phi_R + 2 \Omega_R (t-\tau) \big) \nonumber \\
		& + \frac{32}{15} \nu A_R R_R^4 \Omega_R^7 (t-\tau) \sin \big( \Phi_R + \Omega_R (t-\tau) \big)  + \delta_\Omega^{v^{10}}
\end{align}
Notice, again, that in adding these contributions to (\ref{drho}) we encounter non-trivial cancellations.
In particular, the terms proportional to sinusoids cancel exactly, as they must since there is no counter-term of this form. 
The only remaining secular divergence appears in the third term of the first line and is quadratic in $(\tau - t_0)^2$. We identify this term with $\delta _\Omega^{v^{10}}$,
\begin{align}
	\delta _\Omega^{v^{10}} = \frac{ 16\,896 }{ 25 } \nu^2 R_R^{10} \Omega_R^{13} (\tau -t_0)^2
\end{align}
so that we are left with, after scale-setting $\tau = t$,
\begin{align}
\label{drho}
	\omega_{v^{10}}(t) = {} &  \frac{ 904}{15} \nu A_R R_R^4 \Omega_R^6 \cos \Phi_R   - \frac{5}{2} \frac{ A_R^2 \Omega_R}{R_R^2} \cos 2 \Phi_R 
\end{align}

As with the 1-loop calculation, we can calculate the second order contribution to the orbital phase, $\delta \sigma(t)$, by integrating $\delta \rho$ over time,
\begin{align}
	\delta \sigma (t) = {} & - \frac{504}{5} \nu A_B R_B^4 \Omega_B^5 \sin \Phi_B + \frac{ 5 }{4} \frac{ A_B^2 }{ R_B^2 } \sin 2 \Phi_B + \frac{ 504}{5} \nu A_B R_B^4 \Omega_B^5 \sin \big( \Phi_B + \Omega_B (t-t_0) \big) \nonumber \\
		& + \frac{ 5632}{25} \nu^2 R_B^{10} \Omega_B^{13} (t-t_0)^3 - \frac{608}{15} \nu A_B R_B^4 \Omega_B^6 (t-t_0) \cos \big( \Phi_B + \Omega_B (t-t_0) \big) \nonumber \\
		& - \frac{ 96}{5} \nu A_B R_B^4 \Omega_B^7 (t-t_0)^2 \sin \big( \Phi_B + \Omega_B (t-t_0) \big) - \frac{ 5}{4} \frac{ A_B^2 }{R_B^2 } \sin \big( 2 \Phi_B + 2 \Omega_B (t-t_0) \big) 
\end{align}
Proceeding as before we find the $\calO(v^{10})$ contribution to the phase to be
\begin{align}
	\phi_{v^{10}}(t) = {} & \delta_\Phi^{v^{10}} - \frac{504}{5} \nu A_R R_R^4 \Omega_R^5 \sin \Phi_B(t_0) + \frac{ 5 }{4} \frac{ A_R^2 }{ R_R^2 } \sin 2 \Phi_B(t_0) +  \frac{5632}{25} \nu^2 R_R^{10} \Omega_R^{13} (\tau - t_0)^3 \nonumber \\
		& + \frac{5632}{25} \nu^2 R_R^{10} \Omega_R^{13} (t - \tau)^3  - \frac{608}{15} \nu A_R R_R^4 \Omega_R^6 (t-\tau) \cos \big( \Phi_R + \Omega_R (t - \tau) \big) \nonumber \\
		& - \frac{ 96}{5} \nu A_R R_R^4 \Omega_R^7 (t- \tau)^2 \sin \big( \Phi_R + \Omega_R (t-\tau) \big) + \frac{ 504}{5} \nu A_R R_R^4 \Omega_R^5 \sin \big( \Phi_R + \Omega_R (t-\tau) \big) \nonumber \\
		& - \frac{ 5}{4} \frac{ A_R^2 }{R_R^2 } \sin \big( 2 \Phi_R + 2 \Omega_R (t-\tau) \big)
\end{align}
We choose the $\calO(v^{10})$ phase counter term $\delta_\Phi^{v^{10}}$ to cancel the last three terms in the first line of the equation above,
\begin{align}
	\delta_\Phi^{v^{10}} (\tau, t_0) = {} & \frac{504}{5} \nu A_R R_R^4 \Omega_R^5 \sin \Phi_B(t_0) - \frac{ 5 }{4} \frac{ A_R^2 }{ R_R^2 } \sin 2 \Phi_B(t_0) - \frac{5632}{25} \nu^2 R_R^{10} \Omega_R^{13} (\tau - t_0)^3
\end{align}
The resulting expression for the $\calO(v^{10})$ phase at $\tau = t$ is then given by
\begin{align}
	\phi_{v^{10}}(t) = {} &  \frac{ 504}{5} \nu A_R R_R^4 \Omega_R^5 \sin \Phi_R - \frac{ 5}{4} \frac{ A_R^2 }{R_R^2 } \sin 2 \Phi_R
\end{align}

%--------------------------------------
\subsection{The Renormalization Group solution}
%--------------------------------------

Putting together the order $\varepsilon$ and $\varepsilon^2$ counter-terms we have
\begin{equation}
\begin{aligned}
\label{eq:fullCounterTerms}
	\delta_R = {} & \frac{ 64  \nu }{ 5 }  R_R^6  \Omega_R^6 (\tau-t_0)- \frac{6144}{25}  \nu^2  R_R^{11} \Omega_R ^{12} (\tau-t_0)^2    \\
	\delta_\Omega = {} & - \frac{ 96  \nu }{ 5  } R_R^5 \Omega_R^7(\tau-t_0)+\frac{16896}{25} \nu ^2 R_R^{10} \Omega_R ^{13} (\tau-t_0)^2 \\
	\delta_A = {} & \frac{496}{15} A_R \nu  R_R^5 \Omega_R ^6 (\tau-t_0) \\
	\delta_\Phi = {} & -\Omega_R(\tau-t_0) + \frac{ 48  \nu }{ 5  } R_R^5  \Omega_R^7 (\tau-t_0)^2  - \frac{5632}{25} \nu^2 R_R^{10} \Omega_R^{13} (\tau - t_0)^3  \\
		& + \frac{504}{5} \nu A_R R_R^4 \Omega_R^5 \sin \Phi_B(t_0) - \frac{ 5 }{4} \frac{ A_R^2 }{ R_R^2 } \sin 2 \Phi_B(t_0)
\end{aligned}
\end{equation}
From the expressions relating the bare parameters to the renormalized quantities and counter terms we derive the RG equations through two loops.  The RG equation for $R_R$ through two loops is given by
\begin{align}
	0 = \frac{d}{d\tau}R_B(t_0) = {} &  \frac{dR_R(\tau)}{d\tau} +\frac{ 64 }{ 5 } \nu  R_R^6  \Omega_R^6 + \frac{384 }{5} \nu R_R^5 \Omega_R^5 (\tau - t_0) \bigg( \Omega_R \frac{ d R_R(\tau) }{ d\tau } + R_R \frac{ d \Omega_R (\tau) }{ d\tau } \bigg) \nonumber \\
		& - \frac{12288}{25}  \nu^2  R_R^{11} \Omega_R ^{12} (\tau-t_0) + \calO(v^{15})
\end{align}
At first sight this result seems problematic since it formally diverges.\footnote{In the sense that
	the result depends upon the cut-off, $t_0$.} 
However, solving this equation iteratively in $\varepsilon$ shows that the result is finite (i.e., independent of $\tau - t_0$) leaving
\begin{align}
\frac{dR_R}{d\tau} = - \frac{ 64  \nu }{ 5 }  R_R^6  \Omega_R^6  + \calO (v^{15}).
\end{align}
Similarly, for the orbital angular frequency we have
\begin{align}
	0 =  \frac{ d }{ d\tau } \Omega_B(t_0) = {} & \frac{ d\Omega_R (\tau) }{ d\tau } - \frac{ 96}{5} \nu R_R^5 \Omega_R^7 - \frac{96}{5} \nu R_R^4 \Omega_R^6 (\tau - t_0) \bigg( 5 \Omega_R \frac{ d R_R}{d\tau} + 7 R_R \frac{ d\Omega_R }{ d\tau } \bigg)  \nonumber \\
		& + \frac{ 33\,792}{25} \nu^2 R_R^{10} \Omega_R^{13} (\tau - t_0)  + \calO(v^{15})
\end{align}
Again, solving this iteratively we find that the beta function is independent of the regulator and, as in the case with $R_R$, the two-loop correction does not change the beta function, leaving
\begin{align}
	\frac{d\Omega_R}{d\tau} = {} &  \frac{ 96  \nu }{ 5 }  R_R^5  \Omega_R^7  .
\end{align}

Through $\calO(v^{10})$, the beta function for the amplitude of oscillation receives a two-loop correction and induces a nontrivial RG flow described by
\begin{align}
	\frac{d}{d\tau} A_R(\tau)  = -\frac{496}{15} A_R \nu  R_R^5 \Omega_R ^6 ,
\end{align}
which has the solution  
\begin{align}
\label{A}
	A_R(t) = A_R(t_i) \left( \frac{R_R(t) }{ R_R(t_i) } \right)^{31/12} 
		~~ \Longrightarrow ~~
	e_R(t) \equiv \frac{A_R(t)}{R_R(t)} = e_R(t_i) \left(\frac{R_R(t)}{R_R(t_i)}\right)^{19/12} 
\end{align}
where $e_R(t)$ is the $\calO(v_R^5)$ time-dependent eccentricity of the binary's orbit from~\eqref{eq:AtoEcc}. 
The power-law relation between $e_R$ and $R_R$, namely, $R_R \sim e_R^{12/19}$ agrees with the well-known result from Peters~\cite{peters1964gravitational} when $e_R \ll 1$.

Finally, the RG equation through ${\cal O}(v^{10})$ for the phase parameter satisfies
\begin{align}
	0 = \frac{ d}{d\tau } \Phi_B (t_0) = {} & \frac{ d\Phi_R}{d\tau} - \Omega_R - (\tau - t_0) \frac{d\Omega_R }{ d\tau } + \frac{96}{5} \nu R_R^5 \Omega_R^7 (\tau - t_0) + \frac{ 48}{5} \nu R_R^4 \Omega_R^6 (\tau - t_0)^2 \bigg( 5 \Omega_R \frac{ d R_R}{d\tau } + 7 R_R \frac{ d\Omega_R }{ d\tau } \bigg) \nonumber \\
		&  - \frac{16\,896}{25} \nu^2 R_R^{10} \Omega_R^{13} (\tau - t_0)^2  + \calO(v^{15})
\end{align}
Note that the last two terms in~\eqref{eq:fullCounterTerms} for $\delta _\Phi$ do not contribute at this order.
Again, solving this iteratively we find that the beta function is independent of the regulator $t_0$ leaving us with
\begin{align}
	\frac{ d\Phi_R }{ d\tau} = \Omega_R 
\end{align}
Therefore, through 2-loops we see that the RG equations for $R_R$, $\Omega_R$, and $\Phi_R$ are the same as at 1-loop. However, the beta function at 2-loops for the eccentricity receives a nontrivial contribution that induces an RG flow for $e_R$ in time.

Given our solutions to the RG equations we may now write down the result for the resummed
orbtial coordinates through $\calO(v^{10})$,
\begin{align}
	r(t) = {} & R_R(t) \bigg( 1 + e_R (t) \sin \Phi_R(t) + \frac{1}{2} e^2_R(t) - \frac{ 29\,696}{75} \nu^2 R_R^{10}(t) \Omega_R^{10}(t) \nonumber \\
		&  \qquad\qquad - \frac{ 656}{15} \nu e_R(t) R_R^5 (t) \Omega_R^5 (t) \cos \Phi_R(t) + \frac{ 1}{2} e^2_R(t) \cos 2 \Phi_R(t) + \calO \big( v_R^{15} \Omega_R (t-t_i) \big)  \bigg) \\
	\omega(t) = {} & \Omega_R(t) \bigg( 1 - 2 e_R(t) \sin \Phi_R(t) + \frac{ 904}{15} \nu e_R(t) R_R^5 (t) \Omega_R^5 (t) \cos \Phi_R(t) - \frac{ 5}{2} e^2_R(t) \cos 2\Phi_R(t) + \calO \big( v_R^{15} \Omega_R (t-t_i) \big)  \bigg) \\
	\phi(t) = {} & \Phi_R(t) + 2 e_R(t) \cos \Phi_R(t) + \frac{ 504}{5} \nu e_R(t) R_R^5 (t) \Omega_R^5 (t) \sin \Phi_R(t) - \frac{ 5}{4} e^2_R(t) \sin 2 \Phi_R(t) + \calO \big( v_R^{15} \Omega_R (t-t_i) \big) 
\end{align}  
where we have included the error terms, which can be derived as discussed in Sec.~\ref{sec:errors}, and have written $A_R = e_R R_R$. 
The expressions for the two-loop renormalized initial conditions are given in \eqref{LOa}-\eqref{LOc} and \eqref{A}. 
It is straightforward to show that these resummed perturbative solutions satisfy the equations of motion through $\calO(v^{10})$ and that $d \phi(t) / dt = \omega(t)$ to the same order.
In the case where the initial data is fine-tuned so as to yield a quasi-circular inspiral (i.e., by setting $A_R(t_i) = e_R(t_i) = 0$), the resummed solutions become
\begin{equation}
\begin{aligned}
	r_{\rm qc} (t) & = R_R(t) - \frac{ 29\,696}{75} \nu^2 R_R^{11}(t) \Omega_R^{10}(t)   \\
	\omega_{\rm qc}(t) & = \Omega_R(t) \\
	\phi_{\rm qc}(t) & = \Phi_R(t)
\end{aligned}
\end{equation}

\begin{figure}
	\includegraphics[width=0.475\columnwidth]{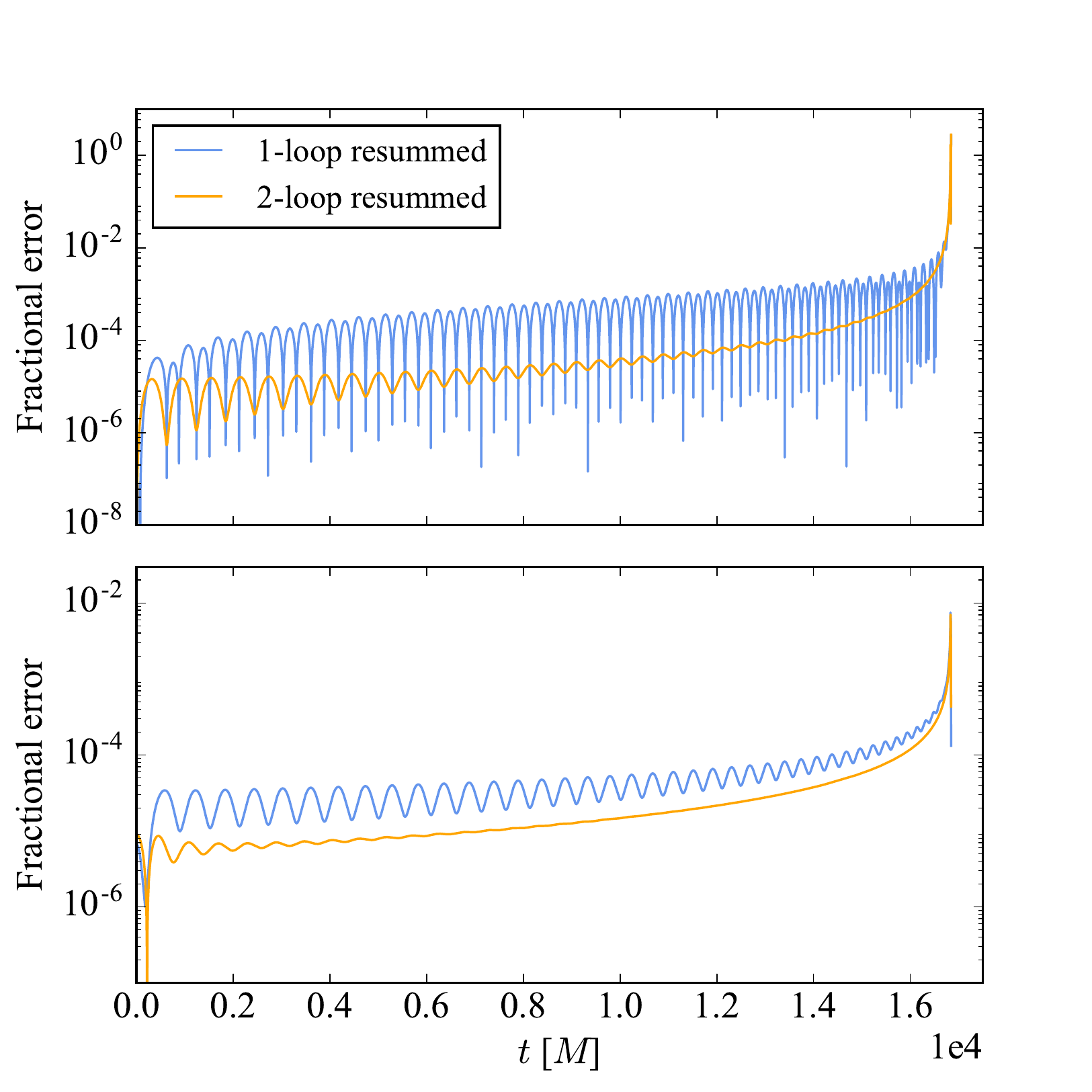}
	\label{fig:comparisonTwoLoops1}
	\caption{Fractional errors for the orbital radius (top panel) and phase (bottom panel)
		between the numerical solution of \eqref{LOBT}, \eqref{eq:initdata} and the one-loop (blue) and two-loop (orange) 
		resummed solutions for the same system and initial conditions shown in Fig.~\ref{fig:comparison1}. }
\end{figure}

Figure~\ref{fig:comparisonTwoLoops1} shows the fractional errors between the numerical solution of \eqref{LOBT} and the one-loop (blue) and two-loop (orange) resummed solutions for the orbital radius (top panel) and phase (bottom panel).
We observe a marked global improvement in the two-loop resummed solution for $r(t)$, providing at least an order of magnitude better accuracy than the one-loop resummed solution.
The two-loop resummed phase shows the same trend as the one-loop solution but is much better at describing the small oscillations due to the $\calO(v^5)$ eccentricity that results from choosing $\dot{r}(0) = 0$ as part of the initial data.

%================================
\section{Beyond Leading Order Radiation Reaction}
%================================

In this formalism the inclusion of higher order radiation reaction forces is straightforward.
The equations of motion through the 1PN correction to radiation reaction forces~\cite{IyerWill:PRL70, IyerWill:PRD52} are given by
\begin{align}
	\ddot{r} - r \omega^2 = {} & - \frac{ M }{ r^2 } + \frac{ 64}{15} \frac{ M^3 \nu }{ r^4 } \dot{r} + \frac{ 16}{5} \frac{ M^2 \nu }{ r^3 } ( \dot{r}^3 + \dot{r} r^2 \omega^2 )  \nonumber \\
		& - \frac{ 8 }{ 105 } \frac{ M^4 \nu }{ r^5 } ( 821 + 210 \nu ) \dot{r} + \frac{ 8 }{ 105 } \frac{ M^3 \nu }{ r^4 } \dot{r} \big( (-362+245 \nu) r^2 \omega^2 - 775 \dot{r}^2 \big) \nonumber \\
		& - \frac{ 4}{35} \frac{ M^2 \nu }{ r^3 } \dot{r} \big( (-65+84 \nu) r^4 \omega^4 + (59+84 \nu) r^2 \omega^2 \dot{r}^2 + 54 \dot{r}^4 \big)
\label{eq:r1PNa}
\end{align}
and
\begin{align}
	r \dot{\omega} + 2 \dot{r} \omega = {} 
		& - \frac{ 24 }{5} \frac{ M^3 \nu }{ r^3 } \omega - \frac{ 8 }{5} \frac{ M^2 \nu }{ r^2 } \omega (\dot{r}^2 + r^2 \omega^2 )  \nonumber \\
		& + \frac{ 4}{105} \frac{ M^4 \nu }{ r^4 } (1325 + 546 \nu ) \omega - \frac{ 2 }{105} \frac{ M^3 \nu }{ r^3 } \omega \big( 2 (205+777\nu) r^2 \omega^2 - (1025 + 1414 \nu) \dot{r}^2 \big) \nonumber \\
		& + \frac{2}{35} \frac{ M^2 \nu }{ r^2 } \omega \big( (313 + 42 \nu) r^4 \omega^4 - (1747 - 42\nu) r^2 \omega^2 \dot{r}^2 + 40 \dot{r}^4 \big).
\label{eq:omega1PNa}
\end{align}
We are interested in demonstrating how to handle higher PN order secular terms in DRG so we do not include the 1PN or higher potentials here, which do not (directly) generate secularly diverging perturbations. Of course, a fully consistent orbital solution should include all potentials that contribute to a given PN order.

As done in the previous section we expand the solution around the background
including perturbations up to order $v_B^7$. Following (\ref{notation}), where now
$\delta \kappa \sim v_B^7 R_B$ and $\delta \rho \sim v_B^6 \Omega_B \sim v_B^7 / R_B$, 
we find that the perturbed radial and angular frequency solutions contain the following contributions at this order
\begin{align}
	r(t) \supset {} & - \frac{4 \nu}{105} R_R^8 \Omega_R^8(336 \nu-3179)(t-t_0) \\
	\omega(t) \supset {} &  \frac{2 \nu}{35} R_R^7 \Omega_R^9(336 \nu-3179)(t-t_0)  \\
	\phi(t) \supset {} & \frac{1}{35} \nu R_R^7 \Omega_R^9(336 \nu-3179)(t-t_0)^2
\end{align}
At this order there is no mixing between
the subleading corrections (i.e., $\delta \rho\delta \kappa$) and the leading $v^5$ 
pieces  ($\delta r$ and $\delta \omega$). As such, there are no quadratic divergences
in $r(t)$ and $\omega(t)$. 

The associated counter-terms then lead to the following RG equations
\begin{align}
	\frac{d R_R}{d\tau}(\tau) = {} & - \frac{ 64  \nu }{ 5 }  R_R^6  \Omega_R^6-\frac{4\nu}{105}  (336 \nu-3179) R_R^8 \Omega_R^8   \\
	\frac{d \Omega_R}{d\tau}(\tau) = {} & \frac{ 96  \nu }{ 5 }  R_R^5  \Omega_R^7+\frac{2\nu}{35} (336 \nu-3179)  R_R^7 \Omega_R^9  \\
	\frac{d \Phi_R}{d\tau}(\tau) = {} & \Omega_R 
\end{align}
The exact solutions to the frequency and phase RG equations are
\begin{align}
	\Omega_R(t) = {} & \Omega_R(t_i) \bigg( \frac{R_R(t)}{R_R(t_i)} \bigg)^{3/2} = \frac{ M^{1/2} }{ R_R^{3/2}(t)  }  \\
	- \frac{32 \nu}{5} M^{5/2} \big( \Phi_R(t) - \Phi_R(t_i) \big) = {} & \frac{1}{5} \big( R_R^{5/2}(t) - R_R^{5/2}(t_i) \big) + \frac{1}{3} \alpha M  \big( R_R^{3/2}(t) - R_R^{3/2}(t_i) \big) + \alpha^2 M^2 \big( R_R^{1/2}(t) - R_R^{1/2}(t_i) \big) \nonumber \\
		& - \alpha^{5/2} M^{5/2} \left[ \tanh ^{-1}  \sqrt{ \frac{ R_R(t) }{ \alpha M} } - \tanh ^{-1} \sqrt{ \frac{ R_R(t_i) }{ \alpha M } } \right] 
\end{align}
where 
\begin{align}
	\alpha \equiv \frac{3179}{336} - \nu  \approx 9.5 - \nu.
\end{align}
and we have used the fact that the combination $R_R^3 \Omega_R^2=M$ is an RG invariant.
Here and below we choose the RG scale $\tau$ to be the observation time $t$.

The solution to the radial RG equation is found by first writing it as
\begin{align}
	\frac{ R_R^4 }{ R_R - \alpha M  } \, dR_R = {} & - \frac{ 64}{5} M^3 \nu \, d\tau.
\end{align}
Integrating both sides gives the exact but implicit relation,
\begin{align}
	-\frac{64 \nu}{5} M^3 (t -t_i) = {} & \frac{1}{4} \big( R_R^4 (t) - R_R^4(t_i) \big) + \frac{1}{3} \alpha M \big( R_R^3(t) - R_R^3 (t_i) \big) + \frac{1}{2} \alpha^2 M^2 \big( R_R^2(t) - R_R^2(t_i) \big) \nonumber \\
		& + \alpha^3 M^3 \big( R_R(t) - R_R(t_i) \big) + \alpha^4 M^4 \log\bigg( \frac{ R_R(t) - \alpha M }{ R_R(t_i) - \alpha M } \bigg)  .
\end{align}
Note that setting $\alpha = 0$ in these RG solutions recovers the one-loop 0PN results derived in the previous sections.
The RG solution for the one-loop oscillation amplitude is also given exactly at this order by
\begin{align}
	A_R(t) = A_R(t_i) = {\rm constant}  .
\end{align}
A two-loop calculation would induce a nontrivial RG flow for the eccentricity as in the 0PN example in the previous section.

%================================
\section{Conclusion}
%================================

In this work we have showed how to utilize the dynamical renormalization group formalism to solve for the long-time behavior for binary inspirals by systematically resumming secularly growing perturbations.
By utilizing this formalism one can avoid the ambiguities intrinsic to using  the adiabatic approximation and orbit-averaging~\cite{pound2008osculating, pound2005limitations}. 
We generated an analytic form for the trajectory of an inspiral at second order in the leading (2.5PN) radiation reaction force.
At this order there exist highly non-trivial consistency checks of the formalism.
In particular, it must be that all secularly divergences have the right functional form to be absorbable into the the initial conditions for the orbit. 
This attribute is called ``renormalizability" in the context of field theory.
Since this formalism solves the equations of motion directly (i.e., without appealing to any kind of averaging procedure) then to go beyond 1PN accuracy one would require the 2PN correction to the radiation reaction force.

Perhaps the most fertile ground for this formalism is  in spin dynamics, where finding closed form solutions becomes a significant challenge.
In nearly all studies of spin effects on compact binary inspiral evolutions, the equations describing the components of the spin vectors are orbit-averaged and, more recently, precession-averaged~\cite{Kesden:2014sla, Gerosa:2015tea}.
However, the dynamical renormalization group does not require averaging over short time scales in the problem to render the problem more amenable for solving, whether analytically or numerically. 
Instead, a naive perturbative solution of the full, non-averaged equations of motion for the binary's orbital coordinates and spin vectors is the starting point for the dynamical renormalization group method.
Incorporating spin effects will be the subject of a companion paper~\cite{spinDRG}.

We thank Luc Blanchet, Marc Favata, Bala Iyer, and Nico Yunes for useful discussions and comments on a previous draft. 
C.R.G. was supported by NSF grants CAREER-0956189 and PHY-1404569 to the California Institute of Technology, by the Sherman Fairchild Foundation, and also thanks the Brinson Foundation for partial support. 
I.Z.R. was supported by NSF-1407744.

\appendix

%%%%%%%%%%
% This appendix is commented out but was going to describe 
% the freedom to choose a renormalization scheme.
%%%%%%%%%%
\begin{comment}

%================================
\section{Renormalization scheme dependence at two-loops}
\label{app:scheme}
%================================

\cg{Note: This isn't complete yet, just throwing these equations as a placeholder for now. This appendix isn't necessary but might help round out the story at 2-loops.}

\begin{equation}
\begin{aligned}
	A_B \to {} & A_B - \frac{ 15}{4} \frac{ A_B^2 }{ R_B } \sin \Phi_B + \frac{ 29\,696}{75 } \nu^2 R_B^{11} \Omega_B^{10} \sin \Phi_B + \frac{ 32}{3} \nu A_B R_B^5 \Omega_B^5 \sin 2 \Phi_B + \frac{ 5}{4} \frac{ A_B^2 }{ R_B } \sin 3 \Phi_B  \\
	\Phi_B \to {} & \Phi_B + \frac{3}{4} \frac{ A_B}{R_B} \cos \Phi_B + \frac{ 29\,696}{75} \frac{ \nu^2 R_B^{11} \Omega_B^{10} }{ A_B } \cos \Phi_B + \frac{32}{3} \nu R_B^5 \Omega_B^5 \cos 2 \Phi_B + \frac{ 5 }{ 4} \frac{ A_B }{R_B} \cos 3 \Phi_B \\
	R_B \to {} & R_B +\mu - \frac{272}{5} \nu A_B R_B^5 \Omega_B^5 \cos \Phi_B - \frac{ 3 A_B^2 }{R_B} \cos 2 \Phi_B  \\
	\Omega_B \to {} & \Omega_B - \frac{3}{2} \frac{  \Omega_B}{R_B} \mu + \frac{408}{5} \nu A_B R_B^4 \Omega_B^6 \cos \Phi_B + \frac{9}{2} \frac{ A_B^2 \Omega_B}{R_B} \cos 2 \Phi_B
\end{aligned}
\end{equation}

\end{comment}
%%%%%%%%%%
%%%%%%%%%%

\bibliography{draft}

\end{document}